\documentstyle[aps]{revtex}
\def\lsim{\mathrel{\rlap{
\lower4pt\hbox{\hskip-3pt$\sim$}}
    \raise1pt\hbox{$<$}}}     
\def\gsim{\mathrel{\rlap{
\lower4pt\hbox{\hskip1pt$\sim$}}
    \raise1pt\hbox{$>$}}}     

\def\beq{\begin{eqnarray}}
\def\eeq{\end{eqnarray}}
\def\bitem{\begin{itemize}}
\def\eitem{\end{itemize}}
\newcommand{\be}{\begin{eqnarray}}
\newcommand{\ee}{\end{eqnarray}}

\def\np{Nucl. Phys.}
\def\pr{Phys. Rev.}
\def\prl{Phys. Rev. Lett.}
\def\pl{Phys. Lett.}

\def\Tr{{\rm Tr}}
\def\L{{\cal L}}

\def\la{\langle}
\def\ra{\rangle}
\def\der{\mbox{d}}

\begin{document}                                                              
\draft
\title{QCD VACUUM CHANGES IN NUCLEI
\footnote{Based on invited talks given at the International Workshop on
Hadrons in Dense Matter, July 3-5, 1996, GSI, Darmstadt, Germany and
at the International
Symposium on Non-Nucleonic Degrees of Freedom Detected in Nucleus, 
September 2-5, 1996, Osaka, Japan}
}
\vskip 0.8cm
\author{Mannque Rho}
\address{
C.E.A.-- Saclay, Service de Physique Th\'eorique,  F-91191 Gif-sur-Yvette 
Cedex, France
\\
{\rm E-mail: rho@spht.saclay.cea.fr}}
\maketitle
\begin{abstract}

In this talk, I discuss how the changes in the QCD vacuum induced by increasing
nuclear matter density affect nuclear properties under normal as well as 
extreme conditions. The quark condensate which is the order parameter for 
the mode in which chiral symmetry is manifested is expected to change as 
matter density or temperature is changed.
The topics discussed are ``BR scaling," its connection
to the structure of nuclei in Landau's Fermi liquid theory and 
a variety of consequences on such nuclear properties as  effective nucleon
mass, nuclear gyromagnetic ratios, $g_A^\star$, axial-charge transitions in
nuclei and on the fluctuations of nuclear matter into the strange-flavor 
direction as observed in heavy-ion collisions. I will also present a simple 
explanation of the recent dilepton data in the CERN-CERES heavy-ion 
experiments involving densities greater than that of nuclear matter and of 
the Indiana results on the longitudinally polarized proton scattering from 
heavy nuclei, evidencing dropping vector-meson masses in medium which can 
be interpreted as signaling an aspect of chiral symmetry
restoration in dense medium. The test of BR scaling provides a bridge 
between the physics of extreme conditions (e.g. relativistic heavy-ion 
collisions) and the physics of normal conditions which has conventional 
descriptions, thereby setting the stage for 
formulating many-body theory for nuclear matter starting from an effective 
chiral Lagrangian.
\end{abstract}

\vskip 1cm

\section{Introduction}
\indent
In the presently accepted theory of strong interactions, the observed
masses of low-energy particles such as
the vector mesons $\rho$ and $\omega$, the baryons $N$, $\Delta \cdots$ 
that make up the ingredient for nuclei and nuclear matter are understood to
originate {\it mainly}
from the spontaneous breaking of chiral $SU(N_f)\times SU(N_f)$
(where $N_f$ is the number of light-quark flavors) to it diagonal subgroup
$SU(N_f)$ characterized by the condensate
\be
\la 0|\bar{q}q|0\ra
\ee
where $|0\ra$ denotes the zero baryon density vacuum. Suppose one has a 
chunk of nuclear matter at a density $\rho_0$ and temperature $T$.
In such an extended system of nuclear matter there 
is a good reason to believe as recent
calculations (to be described below) indicate that the quark condensate will
be modified such that its magnitude decreases as density and/or temperature
is increased. We expect in general that
\be
|\la 0^\star|\bar{q}q|0^\star\ra| < |\la 0|\bar{q}q|0\ra| \ \ \ \
{\rm for}\ \ \ \ \rho >0 
\ee  
where $0^\star$ stands for the ``vacuum" in the presence of the medium.
Now it is expected (although there is no rigorous proof) this condensate will
vanish at certain critical density $\rho_c$ or temperature $T_c$. 

An immediate and exciting question is then how the masses of the hadrons 
living in medium get affected by this change of the condensate and how
the vacuum change affects -- and manifests itself in -- nuclear properties
measured in the laboratory as density $\rho$ and/or temperature $T$
is increased. Can ordinary nuclear phenomena be expressed in terms of the vacuum 
change manifested in the condensate change? How do the properties of hadrons
change as the quark condensate dissolves at the phase transition? Can many-body 
dynamics of nuclei be simplified if expressed in terms of the QCD vacuum
variable ? Indeed it has been argued recently that hadron masses of the light-quark 
constituent will decrease gradually -- and not abruptly -- to nearly zero at the critical
point \cite{br91,bbr96}. These and related questions will occupy much of the
future experimentations at CERN. RHIC and CEBAF.

In what follows, I would like to discuss a recent theoretical work done in
answering the questions posed above. It will be based on ``BR scaling" 
proposed by Gerry Brown and myself in \cite{br91}.
\section{BR Scaling}
\indent
In \cite{br91}, it was argued that nuclear properties can be simply and 
concisely described in tree order with an effective chiral Lagrangian with
the parameters of the Lagrangian given by the scaling
\be
\frac{m_M^\star}{m_M}\approx \sqrt{\frac{g_A}{g_A^\star}}
\frac{m_B^\star}{m_B}\approx \frac{f_\pi^\star}{f_\pi}\equiv
\Phi (\rho)\label{br}
\ee
where the subscripts $M$ and $B$ stand for chiral-quark mesons and 
baryons, respectively, and the star denotes quantities in medium.
The BR scaling relation (\ref{br}) that relates the dropping of light-quark
non-Goldstone-boson masses to that of the nucleon mass which in
turn is related to that of the pion decay constant was first derived
by incorporating the trace anomaly of QCD into an effective chiral Lagrangian. 
The basic idea is that there are two components in the scalar field that 
interpolates the dimension-four operator $\Tr\ G_{\mu\nu} G^{\mu\nu}$, 
one the higher energy gluonium component $G$ and the other lower-energy 
quarkonium component $S$ and that the higher component is integrated out 
from the action. The resulting effective chiral action can then be written 
with the $S$ field coupled in to the Goldstone boson fields and light-quark 
matter fields. We assume that at a given background with a density $\rho$ 
(whether in equilibrium  or not), we can rewrite the effective Lagrangian 
shifted around the background defined by the ``vacuum" expectation value 
$\la S\ra_\rho$ in the way both chiral invariance and scale invariance are 
preserved, with a suitable term that breaks the scale invariance as dictated 
by the trace anomaly of QCD. Symmetries demand that we have a low-energy 
effective chiral Lagrangian of the matter-free space with the parameters of 
the Lagrangian defined as in eq.(\ref{br}). The argument used here is quite 
heuristic but one can obtain the same results starting with a chiral 
Lagrangian consisting of Goldstone bosons, baryons (as well as vetor bosons) 
as matter fields and multi-Fermi terms appropriate for many-nucleon systems 
\cite{br95,brPR95,pmr96} with the four-Fermi interactions accounting for
both the vector 
and scalar interactions needed in Walecka theory \cite{waleckamodel,gelmini}. 
Multi-Fermi interactions in the scalar channel give rise to BR scaling.

\section{BR Conjecture and Landau Fermi-Liquid Theory}
\indent
A contact of the present theory with many-body theory of nuclear matter can be
made through the
reinterpretation of the BR scaling in terms of a
baryon chiral Lagrangian in the relativistic baryon
formalism. Consider 
the Lagrangian containing the usual pionic piece $\L_\pi$, the pion-baryon
interaction $\L_{N\pi}$ and the four-Fermi contact interactions
\be
\L_{4}=\sum_\alpha \frac{C_\alpha^2}{2} (\bar{N}\Gamma_\alpha N)
(\bar{N}\Gamma^\alpha N)
\ee
where the
$\Gamma^\alpha$'s are Lorentz covariant quantities -- including derivatives --
that have the
correct chiral properties. The leading chiral order  four-Fermi contact
interactions relevant for the scaling masses are of the form
\be
\L_{4}^{(\delta)}
=\frac{C_\sigma^2}{2} (\bar{N}N\bar{N}N) -\frac{C_\omega^2}{2}
(\bar{N}\gamma_\mu N\bar{N} \gamma^\mu N).\label{wform}
\ee
As indicated by our choice of notation, the first term can be thought of as
arising when a massive isoscalar scalar meson (say, $\sigma$) is integrated
out and similarly for the second term involving a massive isoscalar vector
meson (say, $\omega$). Consequently, we can make the identification
\be
C_\sigma^2=\frac{g_\sigma^2}{m_\sigma^2},\ \
C_\omega^2=\frac{g_\omega^2}{m_\omega^2}.\label{constant}
\ee
The four-Fermi interaction involving the $\rho$ meson quantum number will be
introduced below, when we consider the  electromagnetic currents.
As is well known \cite{gelmini,br95}, the first four-Fermi  interaction in
(\ref{wform}) shifts the nucleon mass in matter,
\be
m_N^\sigma=m_N-C_\sigma^2 \la \bar{N}N\ra. \label{sigmashift}
\ee
In \cite{br95} it was shown that this shifted nucleon mass scales the same
way as the vector and scalar mesons
\be \frac{m_V^\star}{m_V}\approx
\frac{m_\sigma^\star}{m_\sigma}\approx \frac{m_N^\sigma} {m_N}\approx  \Phi
(\rho).\label{universal}
\ee
This relation was referred to in \cite{br91} as ``universal scaling." There
are two points to note here: First as argued in \cite{br95}, the
vector-meson mass scaling applies also to the masses in (\ref{constant}).
Thus, in medium the meson mass should be replaced by $m_{\sigma,
\omega}^\star$. Consequently, the coupling strengths $C_\sigma$ and
$C_\omega$ are density-dependent.\footnote{I should point out that
for the purpose of the ensuing discussion, neither the detailed knowledge
of the ``heavy" degrees of freedom that give rise to the four-Fermi interactions
nor the specific form of the density dependence will be needed. What really
matters are the quantum numbers involved. The latter is invoked in reducing
various density-dependent parameters to the universal one, $\Phi (\rho)$.}
Second, the scaling can be understood in
terms of effects due to the four-Fermi interactions, which for nucleons on
the Fermi surface correspond to the
fixed-point interactions of Landau Fermi liquid theory according to Shankar
and Polchinski \cite{shankarpol}. We shall establish a direct connection to
the Landau parameters of the quasiparticle interaction.

\subsection{Landau's Effective Mass of the Nucleon}
\indent
The first quantity we will establish is the relation between the nucleon
effective mass $m_N^\star$ and the scaling parameter $\Phi$.
In the Landau-Migdal Fermi liquid theory of nuclear matter
\cite{landau,migdal}, the interaction between two
quasiparticles on the Fermi surface is of the form (neglecting tensor
interactions)
\be
{\cal F}(\vec{p},\vec{p}^\prime) &=& F(\cos \theta) + F^\prime (\cos
\theta)(\vec{\tau}\cdot\vec{\tau}^\prime)\nonumber\\
& +& G(\cos
\theta)(\vec{\sigma}\cdot\vec{\sigma}^\prime)
+ G^\prime (\cos \theta)
(\vec{\tau}\cdot\vec{\tau}^\prime)
(\vec{\sigma}\cdot\vec{\sigma}^\prime),
\eeq
where $\theta$ is the angle between $\vec{p}$ and
$\vec{p}^\prime$. The function $F(\cos \theta)$ can be expanded in
Legendre polynomials,
\beq
F(\cos \theta) = \sum_l F_l P_l(\cos \theta),
\eeq
with analogous expansions for the spin- and isospin-dependent
interactions. The coefficients $F_l$ etc. are the Landau Fermi liquid
parameters. Some of the parameters can be related to physical
properties of the system. The relation between the effective mass and
the Landau parameter $F_1$ (eq.~(\ref{mstar})) is crucial for our
discussion. In what follows, we will be concerned with the spin-independent
Landau parameters, although the $g_A^\star$ problem could be described in 
terms of the spin-isospin-dependent parameter $G^\prime$.

An important point of this paper is that one must distinguish between
the effective mass $m_N^\sigma$, which is of the same form as
Walecka's effective mass, and the Landau effective mass, which
is more directly related to nuclear observables.
To see what the precise relation is, we include the non-local
four-Fermi interaction due to the one-pion exchange term, $L^{(\pi)}_4$.

The total four-Fermi interaction that enters in the renormalization-group
flow consideration \`a la Shankar-Polchinski is then the sum
\be
\label{4fermii}
\L_4=\L_4^{(\pi)} +\L_4^{(\delta)}.
\ee
The point here is that the non-local one-pion-exchange term brings
additional contributions to the effective nucleon mass on top of the
universal scaling mass discussed above.
\noindent
We now compute the nucleon effective mass with the chiral Lagrangian and
make contact with the results of Fermi liquid theory \cite{frimanrho,FR}.
We start with the single-nucleon energy in the non-relativistic
approximation\footnote{We treat the scalar and vector fields
self-consistently and the self-energy from
the pion exchange graph as a perturbation.}
\be
\epsilon (p) =\frac{p^2}{2 m_N^\sigma} +
C_\omega^2\la N^\dagger N\ra +\Sigma_\pi (p)
\label{energy}
\ee
where $\Sigma_\pi (p)$ is the self-energy from the pion-exchange Fock
term.
The self-energy contribution from the vector meson (second term on the
right hand side of (\ref{energy})) comes from an $\omega$ tadpole 
(or Hartree) graph.
The Landau effective mass $m_L^\star$ is related to the quasiparticle
velocity at the Fermi surface
\be
\label{velo}
\frac{\der}{\der p} \epsilon (p)|_{p=p_F}=\frac{p_F}{m_L^\star}
= \frac{p_F}{m_N^\sigma} +\frac{\der}{\der p}\Sigma_\pi (p)|_{p=p_F}.
\ee
Using Galilean invariance, Landau \cite{landau} derived a relation between the
effective mass of the quasi-particles and the velocity dependence of the
effective interaction described by the Fermi-liquid parameter $F_1$:
\be
\label{mstar}
\frac{m^\star_L}{m_N} = 1 + \frac{F_1}{3} = (1-\frac{\tilde{F_1}}{3})^{-1},
\ee
where $ \tilde{F_1} = (m_N/m^\star_L) F_1$.
The corresponding relation
for relativistic systems follows from Lorentz invariance and has been
derived by Baym and Chin \cite{baymchin}.


With the four-Fermi interaction (\ref{4fermii}), there are two distinct
velocity-dependent terms in the quasiparticle interaction, namely the
spatial part of the current-current interaction and the exchange (or Fock)
term of the one-pion-exchange. In the nonrelativistic approximation, their
contributions to $\tilde{F_1}$ are ($\tilde{F_1} = \tilde{F_1^\omega} +
\tilde{F_1^\pi}$) 
\be
\label{fomega}
\tilde{F_1^\omega}&=&\frac{m_N}{m_L^\star} F_1^\omega=
-C_\omega^2\frac{2p_F^3}{\pi^2 m_N^\sigma },\\
\tilde{F_1^\pi}&=& -3\frac{m_N}{p_F}\frac{\der}{\der p}\Sigma_\pi (p)|_{p=p_F},
\ee
respectively. 

Using eq.~(\ref{velo}) we find
\be
(\frac{m_L^\star}{m_N})^{-1}
=\frac{m_N}{m_N^\sigma} +\frac{m_N}{p_F}\frac{\der}{\der p}\Sigma_\pi
(p)|_{p=p_F} = 1-\frac 13 \tilde{F_1}\label{LandauM},
\ee
which implies that
\be
\frac{m_N}{m_N^\sigma}=1-\frac 13 \tilde{F}_1^\omega
=\Phi^{-1}.\label{Phidefined}
\ee
This formula which relates a part of the effective nucleon mass to 
the universal scaling factor $\Phi$ gives a relation between 
the $\sigma$-nucleon interaction
(eq.~(\ref{sigmashift})) and the $\omega$-nucleon coupling 
(eq.~(\ref{fomega})). The $\omega$-exchange contribution to the Landau
parameter $F_1$ is due to the velocity-dependent part of the
potential, $\sim \vec{p}_1\cdot\vec{p}_2/m_N^2$.
This is an ${\cal O}(p^2)$ term, and consequently suppressed in the 
naive chiral
counting.  Nonetheless it is this chirally non-leading term in the
four-Fermi interaction (\ref{wform}) that appears on the same footing
with the chirally leading terms in the $\omega$ and $\sigma$ tadpole
graphs. This shows that there must be subtlety in the chiral counting in the
presence of a Fermi sea.

The pion contribution to $F_1$ can be evaluated explicitly \cite{br80}
\be
\frac 13
\tilde{F_1^\pi}&=& -\frac{3f_{\pi NN}^2m_N}{8\pi^2p_F}[\frac{m_\pi^2+2p_F^2}
{2p_F^2} \ln\frac{m_\pi^2+4p_F^2}{m_\pi^2}-2]\nonumber\\
&\approx& -0.153.\label{FockM}
\ee
Here $f_{\pi NN}\approx 1$ is the non-relativistic $\pi$N coupling
constant.
The numerical value of $\tilde{F_1^\pi}$ is obtained at nuclear matter density,
where $p_F\approx 2m_\pi$.

One of the important results of this paper is that
eq.~(\ref{Phidefined}) relates the only unknown
parameter $\tilde{F}_1^\omega$ to the universal scaling factor $\Phi$.
Note that in the absence of the one-pion-exchange interaction -- and
in the nonrelativistic approximation --
$m_N^\sigma$ can be identified with the Landau effective mass
$m_L^\star$. In its presence,
however, the two masses are different due to the pionic Fock term.
We propose to identify the scaling nucleon mass defined in
eq.~(\ref{br}) with the Landau effective mass:
\be\label{mrelation}
m_L^\star=m_N^\star.
\ee
We note that the Landau mass is defined at the Fermi surface, while the
scaling mass refers to a nucleon propagating in a ``vacuum" modified
by the nuclear medium. Although  
the two definitions are closely related, their precise
connection is not understood at present. Nevertheless,
eq.~(\ref{mrelation}) is expected to be a good approximation. 

\subsection{Axial-Charge Transitions in Nuclei}
\indent
One early evidence for the scaling factor $\Phi$
was found in nuclear axial-charge transitions \cite{denys,warburton}:
\be
A(J^+)\leftrightarrow B(J^-) \ \ \ \ \Delta T=1 \label{axialcharge}.
\ee
This process has been extensively studied recently, experimentally by Warburton
\cite{warburton} and theoretically by several authors \cite{kr,pmr93}.
It has confirmed the prominent role of the pionic degree of
freedom in nuclei as predicted sometime ago in \cite{KDR} and recently 
given a strong theoretical support in terms of chiral perturbation theory
\cite{pmr93}.

The process (\ref{axialcharge}) is dominated by the time component of the 
weak axial current $A_0$ and the axial charge operator relevant to nuclear
processes is given to a very good accuracy by the single-particle current
and a two-body soft-pion-exchange current with three - or higher-body
contributions strongly suppressed \cite{KDR,pmr93}, both of
which goes as \footnote{This behavior for the two-body soft-pion exchange 
current 
follows from the observation that the properties of the pion do not get modified
in medium as one sees in nature.}
\be
A_0 \sim f_\pi^{-1}.
\ee
This shows that in medium the current will be modified by a factor 
$\frac{f_\pi^\star}{f_\pi}\equiv \Phi$ compared with the free-space operator.
Now if one denotes by $R$ the ratio of the one-body matrix element to 
the two-body
matrix element taken between the initial and final nuclear states
\be
R=\la B|A_0^{(2body)}|A\ra/\la B|A_0^{(1body)}|A\ra\label{R}
\ee
then the ratio studied by Warburton
\be
\epsilon_{MEC}\equiv \frac{\la B|A_0|A\ra}{\la B|A_0^{imp}|A\ra}
\ee
where $A_0^{imp}$ stands for the impulse-approximation operator with the 
free-space $f_\pi$, is simply given by
\be
\epsilon_{MEC}=\Phi^{-1} (1+R).\label{epsilon}
\ee
Because of the soft-pion dominance in the two-body charge operator as found
in \cite{KDR}, the ratio $R$ is large, so one sees that the $\epsilon_{MEC}$
can be rather large. The quantity $\epsilon_{MEC}$ is then a direct measure 
of one  part of the relation (\ref{br}).
\subsection{Effective Gamow-Teller Constant $g_A^\star$}
\indent
The renormalized axial-vector coupling constant $g_A^\star$ can be obtained
by relating the Landau effective mass (\ref{LandauM}) and the effective chiral
mass for the nucleon (\ref{br}):
\be
\frac{g_A^\star}{g_A}=\left(1+\frac 13 F^\pi_1\right)^2=\left(1-\frac 13
\Phi \tilde{F}_1^\pi\right)^{-2}.\label{ga*}
\ee

Empirically the  constant $g_A^\star$
is found to be very near 1 \cite{gAdata}. In the 
past the quenching of the axial constant was explained in terms of the
Landau parameter $g_0^\prime$ operative in the $NN$ -- $\Delta N$ channel
\cite{delta}. Here we obtained a relation that does not invoke an explicitly
 spin-isospin-dependent Landau parameter. 
At present it is not clear what the relation is between
the two formulas, both of which predict about the same value for the constant.

\subsection{Orbital Gyromagnetic Ratios in Nuclei}
\indent
A low-energy observable that relates the ``vacuum" factor $\Phi$ to 
Landau parameters
is the gyromagnetic ratios $g_l^{(p,n)}$ of the proton and the neutron in heavy
nuclei. We first recall the standard
Fermi liquid theory result for the gyromagnetic 
ratio.\footnote{This quantity has been extensively analyzed in terms of
standard exchange currents and their relations, via vector-current
Ward identities, to nuclear forces \cite{riska}.}

\subsubsection{Migdal's formula}
\indent
The response to a slowly-varying electromagnetic field of an odd
nucleon with momentum $\vec{p}$ added to a closed Fermi sea
can, in Landau theory, be represented by the current \cite{migdal,BWBS}
\be
\vec{J}= \frac{\vec{p}}{m_N}\left(\frac{1+\tau_3}{2} +\frac 16
\frac{F_1^\prime -F_1}{1+F_1/3} \tau_3\right) \label{qpcurrent}
\ee
where $m_N$ is the nucleon mass in medium-free space.
The long-wavelength limit of the current is not unique. 
The physically relevant one
corresponds to the limit $q \rightarrow 0, \omega \rightarrow 0$ with
$q/\omega \rightarrow 0$, where $(\omega,q)$ is the four-momentum transfer.
The current (\ref{qpcurrent}) defines the gyromagnetic ratio
\be
g_l=\frac{1+\tau_3}{2} +\delta g_l
\ee
where
\be
\delta g_l=\frac 16 \frac{F_1^\prime -F_1}{1+F_1/3}\tau_3
=\frac 16 (\tilde{F}_1^\prime-\tilde{F}_1)\tau_3.\label{deltalandau}
\ee


\subsubsection{From Chiral Lagrangian}
\indent
Let us first compute the gyromagnetic ratio using the chiral 
Lagrangian and demonstrate that Migdal's result (\ref{deltalandau}) is
reproduced. The derivation will be made in terms of Feynman diagrams.
The single-particle current $\vec{J}_1=\vec{p}/m_N^\sigma$ is
given by a diagram 
with the external nucleon lines dressed by the scalar and vector fields.
Note that it is the universally scaled mass $m_N^\sigma$ that enters,
not the Landau mass. This leads to a gyromagnetic ratio
\be
(g_l)_{sp}=\frac{m_N}{m_N^\sigma}\frac{1+\tau_3}{2}.\label{SP}
\ee
At first glance this result seems to imply the enhancement of the
single quasiparticle gyromagnetic ratio by the factor $1/\Phi$ 
(for $\Phi <1$) over the free space value. However this interpretation,
often made in the literature, is not correct. We have to take into account
the corrections carefully.

The first correction to (\ref{SP})
is the contribution from short-ranged high-energy
isoscalar vibrations corresponding to an $\omega$ meson.
This contribution has been computed by several authors \cite{matsui,suzuki}.
In the nonrelativistic approximation one finds
\be
g_l^\omega=-\frac 16 C_\omega^2\frac{2p_F^3}{\pi^2}
\frac{1}{m_N^\sigma}=\frac 16 \tilde{F}_1^\omega.\label{deltaomega}
\ee
Now using (\ref{Phidefined}), we obtain the second principal result of
this paper,
\be
g_l^\omega=\frac 16 \tilde{F}_1^\omega= \frac 12
(1-\Phi (\rho)^{-1}).\label{f1phi}
\ee
The corresponding contribution with a $\rho$ exchange in the graph
yields an isovector term
\be
g_l^\rho=-\frac 16 C_\rho^2 \frac{2p_F^3}{\pi^2}\frac{1}{m_N^\sigma}
\tau_3=\frac 16 (\tilde{F}_1^\rho)^\prime \tau_3 \label{deltarho}
\ee
where the constant $C_\rho$ is the coupling strength of the four-Fermi
interaction
\be
\delta \L=-\frac{C_\rho^2}{2} (\bar{N}\gamma_\mu \tau^a N
\bar{N}\gamma^\mu \tau^a N).
\ee
In analogy with the isoscalar channel, we may consider this as arising
when the $\rho$ is integrated out from the Lagrangian, and consequently
identify
\be
C_\rho^2=g_\rho^2/m_\rho^2.
\ee
Again in medium, $m_\rho$ should be replaced by $m_\rho^\star$.
The results (\ref{deltaomega}) and (\ref{deltarho})
can be interpreted in the language of chiral perturbation theory
as arising from four-Fermi interaction counterterms in
the presence of electromagnetic
field, with the counter terms saturated by the $\omega$ and $\rho$ mesons
respectively (see eq.~(92) of \cite{pmr96}).


The next correction is the pionic exchange current (known as Miyazawa term) 
which yields \cite{br80}
\be
g_l^\pi = \frac 16 ((\tilde{F}_1^\pi)^\prime -\tilde{F}_1^\pi)\tau_3
=-\frac 29 \tilde{F}_1^\pi \tau_3,\label{glpi}
\ee
where the last equality follows from $(\tilde{F}_1^\pi)^\prime = -(1/3)
\tilde{F}_1^\pi$.
Thus, the sum of all contributions is
\be
g_l&=& \frac{m_N}{m_N^\sigma}\frac{1+\tau_3}{2} +\frac 16
(\tilde{F}_1^\omega +(\tilde{F}_1^\rho)^\prime \tau_3)
+ \frac 16 ((\tilde{F}_1^\pi)^\prime -\tilde{F}_1^\pi)\tau_3\nonumber\\
&=&
\frac{1+\tau_3}{2} + \frac 16 (\tilde{F}_1^\prime -\tilde{F}_1)\tau_3
\label{pred}
\ee
where eq.~(\ref{Phidefined}) was used with
\be
\tilde{F}_1 &=& \tilde{F}_1^\omega+\tilde{F}_1^\pi,\label{ftilde}\\
\tilde{F}_1^\prime &=& (\tilde{F}_1^\pi)^\prime +(\tilde{F}_1^\rho)^\prime.
\label{fptilde}
\ee
Thus, when the corrections are suitably calculated, we do
recover the familiar single-particle gyromagnetic ratio $(1+\tau_3)/2$
and reproduce the Fermi-liquid theory result
for $\delta g_l$ (\ref{deltalandau})
\be
\delta g_l =\frac 16 (\tilde{F}_1^\prime-\tilde{F}_1)\tau_3
\label{deltachiral}
\ee
with $\tilde{F}$ and $\tilde{F}^\prime$ in the theory given entirely by
(\ref{ftilde}) and (\ref{fptilde}), respectively.
Equation (\ref{pred}) shows that the isoscalar gyromagnetic ratio is
not renormalized by the medium 
(other than binding effect implicit in the matrix
elements) while the isovector one is. 
{\it It should be emphasized that contrary to naive expectations,
BR scaling is not in conflict with the observed nuclear
magnetic moments.} We will show below that the theory agrees
quantitatively with experimental data.

%
%
%
%
\section{Testing BR Scaling}
\subsection{$\Phi$ from QCD Sum Rules}
\indent
It is possible to extract the scaling factor $\Phi (\rho)$ from
QCD sum rules -- as well as from an in-medium Gell-Mann-Oakes-Renner
relation  -- and compare with our theory.
In particular, the key information is available from
the calculations of the masses of the $\rho$ meson \cite{sumrules,Jin1}
and the nucleon \cite{CFG,Jin2}
in medium. In their recent
work, Jin and collaborators find (for $\rho=\rho_0$) \cite{Jin1}
\be
\frac{m_\rho^\star}{m_\rho}&=& 0.78\pm 0.08.\label{Jin}
\ee
We identify the $\rho$-meson scaling with the universal
scaling factor,
\be
\Phi (\rho_0)=0.78.\label{sigmaM}
\ee
This is remarkably close to the result that follows from the
GMOR relation in medium \cite{lutz,brPR95,CFG2}
\be
\left(\frac{f_\pi^\star}{f_\pi}\right)^2 (\rho_0)  \approx
\frac{{m_\pi^\star}^2}{m_\pi^2} (1-\frac{\Sigma_{\pi N}\,\rho_0}
{f_\pi^2 m_\pi^2}+\cdots)\approx 0.63,\label{GMOR}
\ee
where the pion-nucleon sigma term $\Sigma_{\pi N}\approx 45$ MeV is used.
In fact, in previous papers by Brown and Rho, the scaling factor
$\Phi$ was inferred from the in-medium GMOR relation. How this scaling factor
behaves as a function of density will have to be determined by heavy-ion 
experiments. First-principle calculations anchored on QCD -- QCD sun rules
or lattice -- are not likely to be forthcoming in the near future.

\subsection{Predictions by Chiral Lagrangian}
\indent
Our theory has only one quantity that is not fixed by the theory, namely
the scaling factor $\Phi (\rho)$ ($\tilde{F}_1^\pi$ is of course
fixed for any density by the chiral Lagrangian.).
Since this is given by the QCD sum rule
for $\rho=\rho_0$, we use this information to make quantitative predictions.
\subsubsection{Effective nucleon mass}
\indent
The Landau effective mass of the nucleon (\ref{LandauM}) is
\be
\frac{m_N^\star}{m_N}&=&\Phi\left(1 + \frac 13 F^\pi_1\right) \nonumber\\
&=& \left(\Phi^{-1}-\frac 13 \tilde{F}_1^\pi\right)^{-1}\nonumber\\
&=& (1/0.78 +0.153)^{-1}=0.69(7)\label{L}
\ee
where we used (\ref{FockM}) and (\ref{sigmaM}). This is in agreement with
the QCD sum-rule result of \cite{Jin2}:
\be
\frac{m_N^\star}{m_N}&=& 0.67\pm 0.05.\label{Jin2}
\ee
The agreement is both surprising and
intriguing since as mentioned above, the Landau mass is ``measured" at
the Fermi momentum $p=p_f$ while
the QCD sum-rule mass is defined in the rest frame, so the direct connection
remains to be established. The prediction (\ref{L}) is also in excellent
agreement with observations in the spectroscopy of heavy nuclei and in
the electron scattering from nuclei.
\subsubsection{Scaling of the vector-meson masses}
\indent
The scaling of the vector meson mass as given by the QCD sum-rule prediction
has recently been discussed in two different contexts.

The first case concerns the recent CERN-CERES experiments on dileptons produced
in the central collision of Si on Au at 200 GeV where an excess delepton 
production was observed around the invariant masses of 300 -- 500 MeV 
\cite{CERES,HELIO}.
This enhanced low-mass 
dilepton production was simply explained by Li, Ko and Brown
\cite{lkb95} in terms of the scaled mass for the $\rho$ meson
\be
{m_\rho^\star}\approx \Phi (\rho) m_\rho.
\ee
Here the scaling factor $\Phi$ is somewhat arbitrarily chosen with its 
normalization fixed at $\rho=0$ and $\rho=\rho_0$. Many other mechanisms for
the dilepton excess have been proposed but up to now, it appears that the
explanation in terms of BR scaling is the only viable explanation. It may 
be possible to arrive at a similar result 
starting with a Lagrangian defined
at zero density and doing dynamical calculations \cite{weise}. Whether or not
the dynamical description will justify the inherently quasiparticle picture
that is implicit in BR scaling is not clear at the moment and will eventually
be clarified by the calculations in progress.

The second case concerns polarized proton scattering from a heavy nucleus.
By separating the longitudinal component of the isovector spin response 
function from the transverse, it has been possible to extract information of
the propagation of the $\rho$ meson in medium. In the 198.5 MeV 
$^{28}$Si ($\vec{p},\vec{p}^\prime$)$^{28}$Si reaction \cite{indiana}, 
the Indiana group
demonstrated that the suppression of 
the partial cross section for spin-longitudinal response can be understood
simply if one assumes that the propagating vector meson has a scaling
behavior consistent with the BR scaling.

\subsubsection{Warburton's $\epsilon_{MEC}$}
\indent
The ratio $R$, (\ref{R}), turns out to be quite insensitive to nuclear models
used and does not depend sensitively on the mass number \cite{KDR}. It comes
out to the next-to-next-to-leading order in chiral perturbation theory 
\cite{pmr93}
to be \footnote{This ratio is dominated as mentioned by the soft-pion exchange
term with the loop corrections amounting to $\sim 10$ \% of this value, so
included in the error estimate.}
\be
R=0.5\pm 0.1
\ee
for medium to heavy nuclei. For the heavy nuclei considered by Warburton
(lead region), we may take $\rho\approx \rho_0$ and hence using $(f_\pi^\star/
f_\pi)^2=\Phi^2 \approx 0.63$ in (\ref{epsilon}), we have
\be
\epsilon_{MEC}=\frac{1}{0.78} (1+0.5\pm 0.1)=1.9\pm 0.1
\ee
which should be compared with Warburton's ``empirical" value
\be
\epsilon_{MEC}^{exp}=1.8 \sim  2.0.
\ee
Here Warburton's range indicates the theoretical uncertainty in the estimate
of the single-particle matrix element that depends on the strength of the
tensor force in nuclei. In fact, BR scaling implies the suppression of the 
tensor force, so the uncertainty will remain unless the single-particle matrix 
element is reevaluated with the BR scaling taken into account.

\subsubsection{Effective axial-vector coupling constant}
\indent
The next quantity of interest is the axial-vector coupling constant
in medium, $g_A^\star$
\be
\frac{g_A^\star}{g_A}=\left(1+\frac 13 F^\pi_1\right)^2=\left(1-\frac 13
\Phi \tilde{F}_1^\pi\right)^{-2},
\ee
which at $\rho=\rho_0$ gives
\be
g_A^\star=1.0(0).
\ee
This agrees well with the observations in heavy nuclei \cite{gAdata}.
Again this is an intriguing result. While it is not understood how this
relation is related to the old one in terms of the Landau-Migdal parameter
$g_0^\prime$ in $NN\leftrightarrow N\Delta$ channel \cite{delta},
it is clearly a short-distance effect in the ``pionic channel"
involving the factor $\Phi$. This is consistent with the argument \cite{pmr93}
that the renormalization of the axial-vector coupling constant in
medium cannot be described in low-order chiral perturbation theory.
\footnote{The $g_A^\star$ calculated here is for a quasiparticle
sitting on top of the Fermi sea and is presumably a fixed-point
quantity as one scales down in the sense of renormalization group flow.
As such, it should be applicable within a configuration
space restricted to near the Fermi surface. I think this is the reason why 
$g_A^\star=1$ was required in the $0\hbar\omega$ Monte Carlo shell-model
calculation of Langanke et al \cite{gAdata}. The consequence of this result
is that if one were to calculate core-polarization contributions
involving multiparticle-multihole configurations mediated by tensor forces,
one should obtain only a minor correction. As Gerry Brown has been arguing for
some time, this can happen because of the suppression of tensor
forces in the presence of BR scaling. Note also that the effective
$g_A^\star$ obtained here has nothing to do with the so-called 
``missing Gamow-Teller strength" often discussed in the literature.} 

\subsubsection{Orbital gyromagnetic ratio}
\indent
The correction to the single-particle gyromagnetic
ratio can be rewritten as
\be
\delta g_l=\frac 49\left[\Phi^{-1} -1 -\frac 12 \tilde{F}_1^\pi\right]\tau_3
\ee
where we  have used (\ref{glpi}) and the assumption that the nonet relation 
$C_\rho^2=C_\omega^2/9$ holds. The nonet assumption would be justified
if the constants $C_\omega$ and $C_\rho$ were saturated by the $\omega$
and $\rho$ mesons, respectively.
At $\rho=\rho_0$, we find
\be
\delta g_l=0.22(7)\tau_3.\label{deltapred}
\ee
This is in agreement with the result \cite{exp} for protons extracted
from the
dipole sum rule in $^{209}$Bi using the Fujita-Hirata relation \cite{FH}:
\be
\delta g_l^{proton}=\kappa/2 = 0.23\pm 0.03.
\ee
Here $\kappa$ is the enhancement factor in the giant dipole sum rule.
Given that this is extracted from the sum rule in the
giant dipole resonance region, this is a bulk property, so our theory
is directly relevant.

Direct comparison with magnetic moment measurements is difficult
since BR scaling is expected to quench the tensor force which is crucial
for the calculation of contributions from high-excitation states
needed to extract the $\delta g_l$. Calculations
with this effect taken into account are not available at present.
Modulo this caveat, our prediction (\ref{deltapred}) compares well
with Yamazaki's analysis \cite{yamazaki} of magnetic moments
in the $^{208}$Pb region
\be
\delta g_l^{proton} &\approx& 0.33,\nonumber\\
\delta g_l^{neutron}&\approx& -0.22
\ee
and also with the result of Arima et al. \cite{arima,yamazaki}
\be
\delta g_l\approx 0.25\tau_3.
\ee
\subsubsection{Kaons in dense matter}
\indent
Given the identification of the background field of the ground-state 
nuclear matter with the mean fields of the chiral Lagrangian with the 
BR-scaled parameters as argued in
\cite{br95}, one can then describe fluctuations around that background in
various flavor directions. Of particular current interest is the kaonic 
excitation
in nuclear matter as well as in dense matter.  Specifically one can ask what
happens to a $K^\pm$ propagating in dense medium as in relativistic heavy-ion 
collisions. One can address this issue in terms of an optical potential felt by the
kaon in medium or equivalently in terms of an effective density-dependent
mass of the kaon. 

It is shown in \cite{br95} that a $K^-$ in nuclear matter feels an effective 
potential given by
\be
S_{K^-} +V_{K^-}\approx \frac 13 (S_N-V_N)
\ee
where $S$ and $V$ stand respectively for scalar and vector potentials and the 
subscripts stand for the particles in the potential. The appearance of the factor
$1/3$ indicates that a constituent quark (or quasiquark) picture has
emerged here. From the phenomenology in
Walecka mean-field theory, we have
\be
(S_N-V_N)\lsim -600\ {\mbox{MeV}}.
\ee
This gives the prediction
\be
S_{K^-}+V_{K^-}\lsim -200\ {\mbox{MeV}}.
\ee
This can be compared with the result of an analysis in $K$-mesic atoms
made by  Friedman, Gal and Batty
\cite{friedman} who find attraction at $\rho\approx 0.97\rho_0$ of
\be
S_{K^-}+V_{K^-}=-200\pm 20\ {\mbox{MeV}}.
\ee
Extended to neutron-rich matter and to higher density, 
one expects that negatively
charged kaons will condense at a matter density between 2 and 3 times 
the normal
matter density \cite{br95}. This means that the effective mass of the kaon 
at that density is comparable to the electron chemical potential in 
compact-star matter.

This is an extremely simple prediction which would be very exciting 
if confirmed.
One cannot take this prediction seriously, however,
unless one can show that corrections to the mean-field result
are small. Indeed, constrained by chiral perturbation theory and the 
ensemble of available data on kaon-nucleon interactions, it does not appear
possible to obtain condensation at this low density
\cite{WRW} unless one adds additional four-Fermi interactions that provide
attraction \cite{lbmr} as seems required for the kaonic atom data.
But these are not unique as (many) such four-Fermi terms cannot yet be 
constrained by experiments. What is significant, however, is that
the presently available data on kaon properties in medium obtained 
by the KaoS and FOPI collaborations at GSI are seen to
provide a non-trivial support to 
this additional attraction \cite{BKL}. More accurate data expected to be
available soon will confirm or refute this theory.

\section{Conclusion}
\indent
The assumptions that underlie the preceding arguments are:
\bitem
\item the BR-scaled chiral Lagrangian in mean field represents, at the 
nuclear matter saturation density $\rho_0$, the Landau Fermi-liquid
fixed point theory; 
\item it is sensible to extrapolate the
BR-scaled Lagrangian to describe, by a simple scaling of the parameters,
processes occurring at any density $\rho\neq \rho_0$;
\item the masses and constants of the effective chiral Lagrangian
scale smoothly all the way to the chiral phase 
transition point.
\eitem
I have  shown that it is possible with the above assumptions 
to link what one observes in relativistic 
heavy-ion collisions that probe densities higher than normal
to what one measures in low-energy experiments. This is suggested to be an
indication that nuclear physics can be described in a continuous way from 
normal to the phase transition point with the vacuum suitably modified
through the scaling parameters. The crucial underlying assumption is that
by shifting the ``vacuum" to that characterized by a given
density, one converts 
an intrinsically strong-coupling theory to a weak-coupling one, thus justifying
the mean-field approximation. 

The scheme described here seems to work remarkably
well. This makes the following unresolved basic questions more poignant:
\bitem
\item How is nuclear matter obtained starting from an effective chiral 
Lagrangian ?
\item How good is the notion of quasiparticles -- for baryons for which
Pauli principle could play a role and for mesons for which no such
mechanism is apparent
-- in a density regime where coupling is strong from the view-point of a
theory defined at zero density?
\item How can one justify the notion of Landau Fermi liquid for nuclear matter
when hyperons can enter the process as in kaon condensation?
\item How can one systematically calculate corrections to the mean-field
approximation with a chiral Lagrangian, the parameters of which are BR-scaled ?
\eitem
Some of these questions are being addressed in a variety of ways 
\cite{weise,bf,wambach}
and will be answered when the experimental data being measured become
available.

\subsection*{Acknowledgments}
\indent
This paper was written while I was visiting Institut f\"ur Theoretische Physik
der Universit\"at M\"unchen as a Humboldtpreistr\"ager. I would like to thank
Wolfram  Weise and the Theory Institute for hospitality and the 
Humboldt Foundation for support.
I am grateful for extensive discussions with Gerry Brown, Bengt Friman,
Chang-Hwan Lee, Chaejun Song and Wolfram Weise.


\begin{references}
\bibitem{br91} G.E. Brown and M. Rho, \prl \ {\bf 66} (1991) 2720.
\bibitem{bbr96} G.E. Brown, M. Buballa and M. Rho, ``A mean-field theory of
the chiral phase transition," nucl-th/9603016, \np, in press.
\bibitem{br95} G.E. Brown and M. Rho, \np \ {\bf A596} (1996) 503.
\bibitem{brPR95} G.E. Brown and M. Rho, Phys. Repts. {\bf 269} (1996) 333.
\bibitem{pmr96} T.-S. Park, D.-P. Min and M. Rho, \np \ {\bf A596} (1996) 515.
\bibitem{gelmini} G. Gelmini and B. Ritzi, \pl \ {\bf B357} (1995) 432.
\bibitem{waleckamodel} B.D. Serot and J.D. Walecka, Ad. Nucl. Phys. {\bf 16}
(1986) 1
\bibitem{shankarpol} R. Shankar, Rev. Mod. Phys. {\bf 66} (1994) 129;
J. Polchinski, in ``Recent Directions in Particle Theory," ed.
J. Harvey and J. Polchinski (World Scientific, 1994)
\bibitem{landau} L.D. Landau, JETP {\bf 3} (1957) 920; JETP {\bf 5} (1957) 101
\bibitem{migdal} A.B. Migdal, {\it Theory of Finite Fermi Systems and
Applications to Finite Nuclei}\ (Interscience, London, 1967)
\bibitem{frimanrho} B. Friman and M. Rho, ``From chiral Lagrangians
to Landau Fermi liquid theory of nuclear amtter," nucl-th/96022025.
\bibitem{FR} B. Friman and M. Rho, to appear
\bibitem{baymchin} G. Baym and S. Chin, \np {\bf A262} (1976) 527
\bibitem{br80} G.E. Brown and M. Rho, \np \ {\bf A338} (1980) 269
\bibitem{denys} D.H. Wilkinson, in {\it Nuclear Physics 
with Heavy Ions and
Mesons},\ ed. R. Balian, M. Rho and G. Ripka (North-Holland, Amsterdam, 1978)
\bibitem{warburton} E.K. Warburton, \prl \ {\bf 66} (1991) 1823.
\bibitem{kr} K. Kubodera and M. Rho, \prl \ {\bf 67} (1991) 3479; I.S. Towner,
\np \ {\bf A542} (1992) 631; M. Kirchbach, D.O. Riska and K. Tsushima, 
\np \ {\bf A542} (1992) 616.
\bibitem{pmr93} T.-S. Park, D.-P. Min and M. Rho, Phys. Repts. {\bf 233}
(1993) 341; T.-S. Park, I.S. Towner and K. Kubodera, \np \ {\bf A579} (1994)
381
\bibitem{KDR} K. Kubodera, J. Delorme and M. Rho, \prl \ {\bf 40} (1978) 755.
\bibitem{gAdata} D.H. Wilkinson, in {\it Nuclear Physics with Heavy Ions and
Mesons},\ ed. R. Balian, M. Rho and G. Ripka (North-Holland, Amsterdam, 1978);
B. Buck and S.M. Perez, \prl \ {\bf 50} (1983) 1975; K. Langanke, D.J. Dean,
P.B. Radha, Y. Alhassid and S.E. Koonin, nucl-th/9504019
\bibitem{delta} M. Rho, \np \ {\bf A231} (1974) 493; K. Ohta and M. Wakamatsu,
\np \ {\bf A234} (1974) 445




\bibitem{riska} K. Tsushima, D.O. Riska and P.G. Blunden, \np \ {\bf
A559} (1993) 543
\bibitem{BWBS} G.E. Brown, W. Weise, G. Baym and J. Speth, Comments Nucl.,Part. Phys. {\bf 17} (1987) 39
\bibitem{matsui} T. Matsui, \np \ {\bf A370} (1981) 365
\bibitem{suzuki} H. Kurasawa and T. Suzuki, \pl \ {\bf B165} (1985) 234
\bibitem{sumrules} T. Hatsuda and S.H. Lee, \pr \ {\bf C46} (1992) R34;
T. Hatsuda, \np \ {\bf A544} (1992) 27c
\bibitem{Jin1} X. Jin and D.B. Leinweber, \pr \ {\bf C52} (1995) 3344,
nucl-th/9510064
\bibitem{CFG} R.J. Furnstahl, D.K. Griegel and T.D. Cohen, \pr \ {\bf C46}
(1992) 1507; T.D. Cohen, R.J. Furnstahl, D.K. Griegel and X. Jin,
Prog. Part. Nucl. Phys. {\bf 35} (1995) 221
\bibitem{Jin2} R.J. Furnstahl, X. Jin and D.B. Leinweber, ``New QCD sum rules
for nucleons in nuclear matter," nucl-th/9511007
\bibitem{CFG2} T.D. Cohen, R.J. Furnstahl and D.K. Griegel, \pr \ {\bf
C45} (1992) 1881
\bibitem{lutz} M. Lutz,  S. Klimpt and W. Weise, \np \ {\bf A542} (1992) 755
\bibitem{CERES} G. Agakichiev et al., \prl \ {\bf 75} (1995) 1272
\bibitem{HELIO} M. Masera et al., \np \ {\bf A590} (1995) 93c
\bibitem{lkb95} G.Q. Li, C.M. Ko and G.E. Brown, \prl \ {\bf 75}
(1995) 4007; nucl-th/9608040, \np \ {\bf A}, in press  and to appear.
\bibitem{weise} F. Klingl and W. Weise, private communication.
\bibitem{indiana} E.J. Stephenson et al, ``The nuclear isovector spin
response in the 198.5-MeV $^{28}$Si ($\vec{p},\vec{p}^\prime$)$^{28}$Si 
reaction," Indiana cyclotron preprint.
\bibitem{exp} R. Nolte, A. Baumann, K.W. Rose and M. Schumacher, \pl
\ {\bf B173} (1986) 388
\bibitem{FH} J.I. Fujita and M. Hirata, \pl \ {\bf B37} (1971) 237
\bibitem{yamazaki} T. Yamazaki, in {\it Mesons in Nuclei},\ ed. M. Rho and
D.H. Wilkinson (North-Holland, Amsterdam, 1979)
\bibitem{arima} A. Arima, G.E. Brown, H. Hyuga and M. Ichimura, \np \ {\bf
A205} (1973) 27
%
%
%
%
%
\bibitem{friedman} E. Friedman, A. Gal and C.J. Batty, \pl \ {\bf B308},
6 (1993); \np \ {\bf A579}, 518 (1994).
\bibitem{WRW} T. Waas, M. Rho and W. Weise, ``Effective kaon mass in 
dense matter: short-range correlations," in preparation.
\bibitem{lbmr} C.-H. Lee, G.E. Brown, D.-P. Min and M. Rho, \np \ {\bf A585}
(1995) 401.
\bibitem{bf} B. Friman, work in progress and to appear.
\bibitem{wambach} G. Chanfray, R. Rapp and J. Wambach, \prl \ {\bf 76}
(1996)368 and work in progress.
\bibitem{BKL} G.E. Brown, C.M. Ko and G.Q. Li, ``From $K^+$ in heavy-ion 
collisions to $K^-$ in kaonic atoms," nucl-th/9608039.






\end{references}
\end{document}